\documentclass[preprint2]{emulateapj}

\usepackage{graphicx}

\shorttitle{Near-Infrared Polarimary of NGC 6334}
\shortauthors{Hashimoto et al.}

\begin{document}
  \title{Wide-Field Infrared Imaging Polarimetry of the NGC 6334 Region:\\
    A Nest of Infrared Reflection Nebulae}
  
  \author{J. Hashimoto\altaffilmark{1}, M. Tamura\altaffilmark{1},
    R. Kandori\altaffilmark{1}, N. Kusakabe\altaffilmark{1},
    Y. Nakajima\altaffilmark{1}, 
    \\ M. Kurita\altaffilmark{2},
    T. Nagata\altaffilmark{3}, T. Nagayama\altaffilmark{3},
    J. Hough\altaffilmark{4},  
    A. Chrysostomou\altaffilmark{5}}
  
  \altaffiltext{1}{National Astronomical Observatory, 2-21-1 Osawa, Mitaka, Tokyo 181-8588; 
    hashmtjn@optik.mtk.nao.ac.jp}
  \altaffiltext{2}{Department of Astrophysics, Nagoya University, Chikusa-ku, Nagoya 464-8602}
  \altaffiltext{3}{Department of Astronomy, Kyoto University, Sakyo-ku, Kyoto 606-8502}
  \altaffiltext{4}{Centre for Astrophysics Research, University of Hertfordshire, Hatfield HERTS AL10 9AB, UK}
  \altaffiltext{5}{Joint Astronomy Centre, 660 N. A'ohoku Place, Hilo, HI 96720}

  \begin{abstract}    
    We report the detection of eighteen infrared reflection nebulae (IRNe)
    in the $J$, $H$, \& $Ks$ linear polarimetric observations of the NGC 6334 massive 
    star-formation complex, of which 16 IRNe are new discoveries.
    Our images cover $\sim$180 square arcminutes, one of the widest near-infrared polarization data
    in star-formation regions so far.
    These IRNe are most likely associated with embedded young OB stars at different evolutionary phases,
    showing a variety of sizes, morphologies, and polarization properties,
    which can be divided into four categories.
    We argue the different nebula characteristics to be a possible evolutionary
    sequence of circumstellar structures around young massive stars. 
  \end{abstract}
  
  \section{Introduction}\label{intro}
  Massive star-formation processes are poorly understood phenomena due to 
  large distances, high extinction, and substantial source confusion.
  Recent infrared observations of (sub)arcsecond high-resolution
  allow one to study the complex nature of OB star-formations. 
  In particular, near-infrared (NIR) imaging linear polarimetry is 
  a powerful tool for investigating the circumstellar structures
  of the young stellar objects (YSOs). The radiation from YSOs still 
  embedded in their parent molecular clouds is often scattered by 
  the dust in their circumstellar structures, and is observed as infrared reflection nebula 
  (IRN). This is also a good morphological tracer of mass outflow \citep{hoda84,sato85},
  especially observed at high resolution \citep{tamu91}.
  In massive star-formation regions, NIR polarimetry
  can distinguish the IRNe from the \ion{H}{2} region nebulae \citep{tamu06},
  which allows one to easily uncover new IRNe buried in various emissions.
  In this Letter, we present wide-field linear polarization of 
  the NGC 6334 region.
  
  NGC 6334 is an active massive star-formation complex,
  a total bolometric luminosity $\sim1.9 \times 10^{6} L_{\odot}$ 
  \citep{loug86}, at a distance of 1.7 kpc \citep{neck78}.
  This giant ($\sim$10 pc) complex has several separated massive star-formation
  sites. \citet{mcbr79} show six strong far-infrared (FIR) sources named NGC 6334 I to VI 
  (see \citet{krae99} for more details on other observations).
  Since each site is in the same parent cloud at the same distance,
  systematic studies of massive star-fomation
  and its evolution are expected to be conducted. 

  \section{Observations \& Data Reduction}\label{obs}
  The NIR linear polarization images were obtained on
  June 25, 30, and July 1 2006 with the 
  SIRIUS camera, Simultaneous three-color InfraRed Imager for 
  Unbiased Survey \citep{naga03} and its polarimeter (SIRPOL; Kandori et al. 2006)
  mounted on the IRSF 1.4 m telescope at 
  the South African Astronomical Observatory in Sutherland.
  The image scale of the array was 0$\farcs$45 pixel$^{-1}$, 
  giving a field of view of $7'.7 \times 7'.7$. 
  The polarizations were measured by stepping the half waveplate to
  four angular positions (0\degr, 45\degr, 22.5\degr and  
  67.5\degr). Ten dithered frames were observed per waveplate position, 
  and we observed 9 sets for the object, giving
  $9 \times 10$ frames of 10 s integration per waveplate position.
  Seeing was 1$\farcs$5 (FWHM) in the $J$ band. 
  
  The data were reduced in the standard manner of infrared image reduction:
  subtracting a dark-frame and dividing by a flat-frame.
  In addition, the data for each waveplate position 
  were registered, and then combined. The Stokes parameters were derived
  in the same manner as \citet{tamu06}.   

  \section{Results \& Discussion}
  The intensity images and linear polarization vector images in the NIR bands
  are shown in Figs.\ref{fig1}(A) to (E). We briefly explain the detected IRNe in \S\ref{3.1}, 
  which are selected from the highly polarized intensity and the polarization vector
  pattern, and then qualitatively discuss their polarization properties in \S\ref{3.2}.

  \subsection{Discovery of Eighteen Infrared Reflection Nebulae}\label{3.1}
  
\begin{figure}
    \epsscale{1}
    \plotone{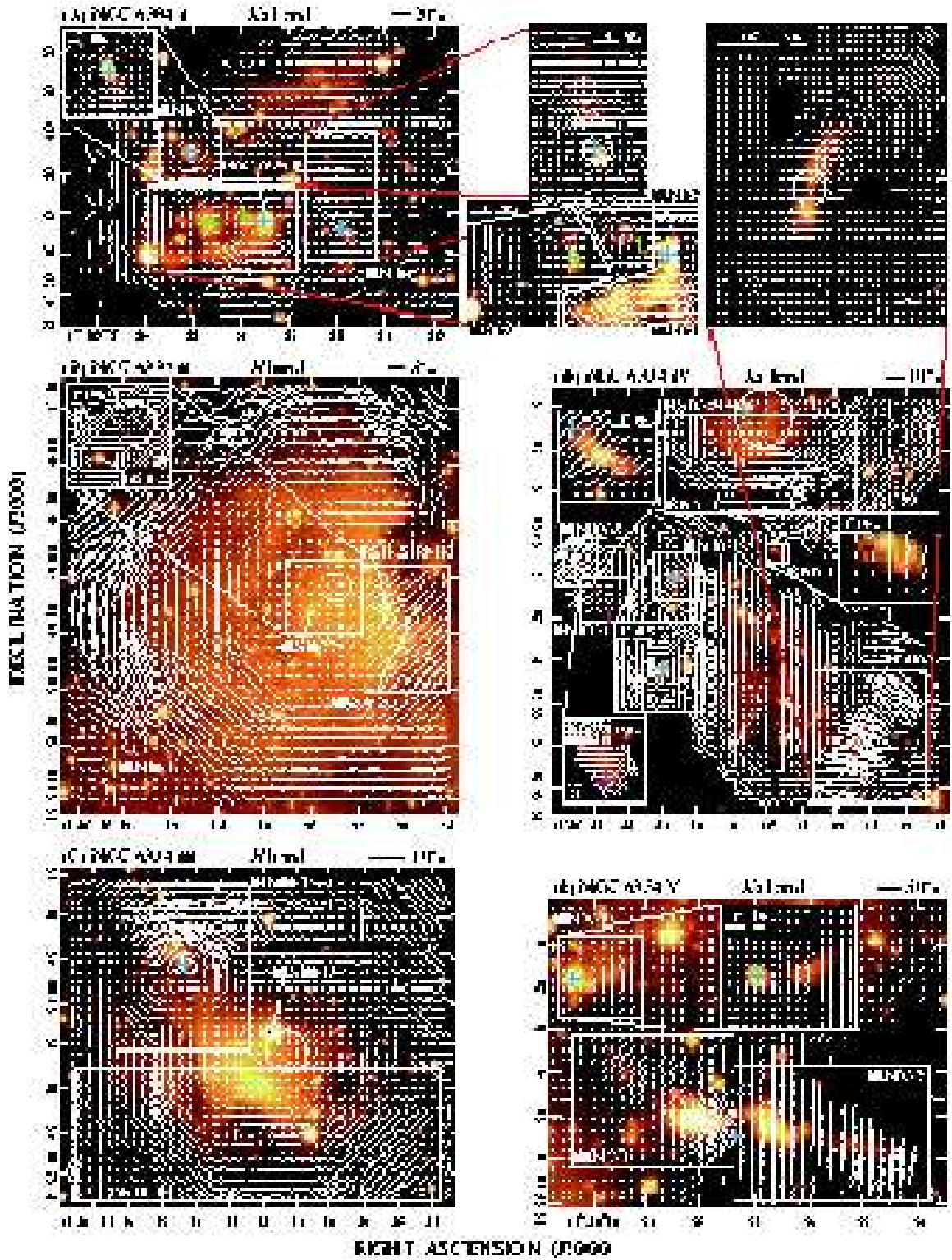}
    \caption{
      Polarization vectors are superposed on
      the intensity images in the $H$ or $Ks$ band.
      In the magnifications, except for IRN IV-6, the vectors are superposed on 
      the polarized intensity images in the same band.
      The white square indicates each IRN,
      and the circle shows each position of the illuminating sources with $1\sigma$ error.
      Blue crosses represent the positions of Spitzer 8.0 $\mu$m sources,
      green crosses represent the MIR sources for IRN I-2 and 3,
      and purple cross represents the radio source for IRN IV-6.
      For IRN I-1, I-5, IV-4, and V-3, since illuminating sources are 
      identified with NIR sources, the error circles are not marked. 
      \label{fig1}}
  \end{figure}

  {\sl NGC 6334 I.} ---
  We identify five IRNe as shown in Fig.\ref{fig1}(A).
  IRN I-1 is illuminated by the ultracompact \ion{H}{2} (UC\ion{H}{2})
  region of NGC 6334 F \citep{rodr82}.
  IRN I-2 and 3 are not illuminated by NGC 6334 F \citep{rodr82} but are
  a monopolar and a bipolar nebula
  illuminated by DPT00 2 \citep{debu02} and KDJ I-3 \citep{krae99},
  respectively.  IRN I-4 is also a bipolar nebula
  associated with an invisible NIR source. IRN I-5 around IRS I-14 \citep{stra89}
  has a concentric polarization.
  
  {\sl NGC 6334 II.} ---
   Fig.\ref{fig1}(B) shows an approximately centrosymmetric polarized pattern.
  Though we calculate the position of the illuminating source
  from the intersection of the normals to the polarization vectors, marked with a 1$\sigma$ error circle 
  in Fig.\ref{fig1}(B), there are no bright sources within the error circle.
  Note that IRS II-22 \citep{stra89} near the error circle is surrounded by
  a local concentric polarization.
  The edge of the spherical \ion{H}{2} region has a higher ($\sim$5 \%) polarization 
  than the vicinity of the error circle ($\sim$1 \%).
  We refer to this region as IRN II-1. Fig.\ref{fig1}(B) also shows that 
  an elongated IRN II-2, clearly seen in a polarized intensity image,
  has a polarization relatively well aligned with $\theta \sim 120\degr$.
  
  {\sl NGC 6334 III.} ---
  Similar to IRN II-1, Fig.\ref{fig1}(C) shows a centrosymmetric polarized pattern 
  on the edge of the \ion{H}{2} region and  the polarization is higher on the edge.
  Since IRS III-13 \citep{stra89} is the brightest NIR source near the error circle,
  we believe it is the illuminating source. We refer to this region as IRN III-1.
  Another IRN is a relatively large bipolar nebula of IRN III-2 associated with 
  a totally invisible source at NIR wavelengths. 

  {\sl NGC 6334 IV.} ---
  Fig.\ref{fig1}(D) shows IRN IV-1, similar to IRN II-1 and III-1, 
  has a centrosymmetric polarized pattern, and  the positional error circle indicates 
  that IRS IV-54 \citep{stra89} is the illuminating source.
  IRN IV-2 and 3, better seen in a polarized intensity image, have relatively well 
  aligned polarizations, $\theta \sim 170\degr$ and $\sim 130\degr$, respectively.
  There is IRN IV-4 with a concentric polarization around IRS IV-10 \citep{stra89}. 
  IRN IV-5 is a bipolar nebula associated with an invisible source at NIR wavelengths.
  The faint IRN IV-6 is associated with an optically thin
  \ion{H}{2} region, G351.24+0.65 \citep{mora90}.
  
  {\sl NGC 6334 V.} ---
  In spite of a bipolar morphology for IRN V-1 and 2 in Fig.\ref{fig1}(E),
  \citet{hash06} showed their illuminating sources are different (see also Chrysostomou et al. 1994).
  IRN V-3 is associated with a monopolar nebula with IRS V-41 \citep{stra89}.
  
  \subsection{Classicifation and Possible Evolution of Infrared Reflection Nebulae}\label{3.2}
  Eighteen IRNe are detected in NGC 6334, and these IRNe can be divided into four categories 
  of Type A to D. We conclude their illuminating sources are embedded or visible OB stars 
  from the previous radio and infrared observations, except IRN I-4, II-1, 2, III-2, IV-3, 
  and 5 without the previous radio and infrared observations (see Table \ref{table1} caption).
  Table \ref{table1} summarizes the list of eighteen IRNe and the nature of their illuminating sources. 

  The first category ({\sl Type A}) contains IRNe with relatively well aligned polarization vectors.
  These IRNe have a non-point-like morphology and extend in a direction relatively perpendicular to 
  polarization vectors. The illuminating sources are invisible at NIR wavelengths.
  The second category ({\sl Type B}) contains monopolar/bipolar IRNe with several ten percent 
  level polarization. The illuminating sources are invisible at NIR wavelengths.
  The third category ({\sl Type C}) contains IRNe with concentric polarization around the visible 
  illuminating sources at NIR wavelengths. These IRNe have a relatively small scale ($\lesssim$0.1 pc). 
  The last category ({\sl Type D}) contains relatively large scale IRNe ($\gtrsim$1 pc).
  The edge of IRNe has centrosymmetric polarization
  and have a higher polarization than the vicinity of the illuminating source.
  
  We argue the above classification can be placed along the approximate evolutionary sequence, i.e., 
  Type A is the youngest and evolve into D through B and C. 

  The reason we conclude Type A is the youngest is that the density of circumstellar
  material seems to be the highest. In this case, in contrast to the case of Type B 
  described below, we expect that originally centrosymmetric polarization in the envelope region 
  can be affected by multiple scattering, resulting in aligned vector patterns 
  as observed (see e.g., Fischer et al. 1994).
  \citet{whit97} in fact detected a similar aligned polarization in
  IRN associated with L1527 in Taurus. This IRN extends in the direction 
  of the outflow \citep{tamu96}, perpendicular to polarization vectors. 
  \citet{zhan07} recently found a compact redshifted 
  outflow, whose extension is well correlated with IRN II-2. 
  We also suggest that the illuminating source of IRN II-2 
  might be invisible at NIR wavelengths, and is therefore not likely 
  to be IRS II-23 or 24. Similarly, the extensions of IRN IV-2 and 3 
  are relatively perpendicular to polarization vectors in each of
  the nebulae. Additionally, these IRNe are located in the vicinity of 
  the submillimeter sources detected by \citet{sand99}.
  Thus we also categorize them as Type A.

  The more evolved Type B contains eight IRNe which are monopolar/bipolar nebulae. 
  Since a close relationship has been suggested between IRNe and CO outflows (Nagata et al. 1984;
  Hodapp et al. 1984; Sato et al. 1985; Yamashita et al. 1989),
  we consider these IRNe correspond to the walls of the outflow material
  where the vectors are nearly centrosymmetric.
  Recently \citet{arce07} summarized the relation between the outflow collimation and its evolution;
  IRN III-2 is categorized as the most evolved one in their classification.

  Small scale Type C contains IRN I-5 and IV-4 associated with 
  the visible NIR sources. The transition from Type B to C we infer is as follows:
  the outflow associated with the Type B sources widens its opening angle 
  with its evolution \citep{arce07},
  then stellar radiation might be scattered by circumstellar material
  more concentric rather than bipolar/monopolar.
  Note that since it is possible that the concentric polarization is 
  due to the inclination of outflow, these IRNe could be categorized as Type B.

  Finally, Type D contains relatively large IRNe.
  The illuminating sources have well-developed \ion{H}{2} regions that
  have had time to expand into the surrounding cloud.
  Due to stronger ionized gas emission in the vicinity 
  of the illuminating sources, the degree of polarization is
  lower near the source than at the edge of the \ion{H}{2} region.
  These IRNe could be due to reflection from dust at the ionization front.

  In Orion, IRNe around IRc2 and BN correspond to Type B,
  and that around the Trapezium \ion{H}{2} region is inferred to be Type D.
  In particular, that of BN is similar to IRN V-3 since
  the monopolar IRN is associated with the NIR visible illuminating source.

  To summarize, we detected eighteen IRNe associated with
  young OB stars and suggest that the difference of the polarized patterns reflects the 
  evolutionary sequence of IRNe; 
  the youngest IRNe with well aligned polarization ({\sl Type A}) evolve into
  monopolar/bipolar IRNe with roughly centrosymmetric polarization ({\sl Type B}). 
  When IRNe have concentric polarizations ({\sl Type C}), 
  the illuminating sources create \ion{H}{2} regions,
  and then, IRNe become less distinct near 
  the central sources with a centrosymmetric circular pattern 
  at the region of the ionization boundary ({\sl Type D}).
  We note, however, that care must be taken to infer above conclusions from the 
  polarization pattern alone.
  Further high resolution radio observations for kinematical information
  should be undertaken to verify our IRN evolutionary hypothesis.

\bigskip  

  We are grateful to an anonymous referee for providing
  many useful comments leading to an improved paper.
  This research was partly supported by Grants-in-Aid for Scientific Research on Priority Areas from 
  the Ministry of Education, Culture, Sports, Science and Technology (MEXT)
  and No.16340061, 16077204, and 16340061.

  \begin{table}
    \begin{center}
      \caption{Infrared reflection nebulae and their illuminating Sources in NGC 6334\label{table1}}
      \begin{tabular}{c|ccccccccc} 
	\tableline
	\tableline
	Type &\multicolumn{2}{c}{IRN} &\multicolumn{7}{c}{Illuminating Source} \\ \cline{4-10} 
	&Name    &Size &\multicolumn{2}{c}{Position (J2000)\tablenotemark{a}} & Identification\tablenotemark{b}&Spitzer& Maser\tablenotemark{c}&$K$ band\tablenotemark{d}&ZAMS Sp.\tablenotemark{e} \\ 
	&        &(pc) &(h: m: s)&\ \ ($\degr:\ ':\ ''$)                  &     &8.0 $\mu$m    && (mag)        &\\
	\hline
	   &IRN II-2   &0.5  & -----     & -----       &invisible                      &No  &---          & ---     &unknown     \\
	 A &IRN IV-2   &0.05 &17 20 17.85& -35 54 53.0 &KDJ IV-3 (MIR), MM2            &No  &---          & ---     &B1          \\
	   &IRN IV-3   &0.1  &17 20 23.82& -35 54 56.3 &MM3?                           &Yes &OH           & ---     &unknown     \\\tableline
	   &IRN I-2    &0.1  &17 20 53.77& -35 47 00.6 &DPT00 2 (MIR)                  &No  &---          & ---     &B4          \\
	   &IRN I-3    &0.2  &17 20 54.59& -35 47 02.6 &KDJ I-3 (MIR), IRS I-9 (NIR)   &No  &---          &12.00    &B0          \\
	   &IRN I-4    &0.2  &17 20 51.90& -35 47 04.2 &invisible                      &Yes &---          & ---     &unknown     \\
	 B &IRN III-2  &0.5  &17 20 34.24& -35 50 47.9 &invisible                      &Yes &---          & ---     &unknown     \\
	   &IRN IV-5   &0.5  &17 20 15.35& -35 56 04.7 &invisible                      &No  &---          & ---     &unknown     \\
	   &IRN V-1    &0.4  &17 19 57.79& -35 57 50.8 &WN-A1 (NIR)                    &No  &NH$_{3}$, OH & ---     &B           \\
           &IRN V-2    &0.4  &17 19 57.46& -35 57 52.5 &KDJ V-4 (MIR)                  &Yes &NH$_{3}$     & ---     &B2          \\
           &IRN V-3    &0.15 &17 20 00.35& -35 57 17.9 &IRS V-41 (NIR)                 &Yes &---          &\ 8.50   &late O      \\\tableline
	 C &IRN I-5    &0.15 &17 20 55.08& -35 46 45.0 &IRS I-14 (NIR)                 &Yes &NH$_{3}$     &11.56    &early B     \\
	   &IRN IV-4   &0.15 &17 20 20.72& -35 55 04.7 &IRS IV-10 (NIR)                &Yes &---          &\ 9.76   &late O      \\\tableline
	   &IRN I-1    &0.3  &17 20 53.50& -35 47 02.8 &NGC 6334 F, IRS I-10 (NIR)     &Yes &OH           &10.24    &B0          \\
	   &IRN II-1   &2.0  &17 20 46.93& -35 49 05.4 &invisible                      &No  &---          & ---     & unknown    \\ 
	 D &IRN III-1  &1.0  &17 20 31.78& -35 51 11.4 &NGC 6334 C, IRS III-13 (NIR)   &No  &---          &\ 7.37   &O8          \\
           &IRN IV-1   &0.6  &17 20 18.62& -35 54 05.1 &IRS IV-54 (NIR)                &Yes &---          &\ 9.87   &late O      \\
	   &IRN IV-6   &0.1  &17 20 22.62& -35 55 19.9 &G351.24+0.65 (radio)           &No  &---          & ---     &B0.5        \\\tableline
	
      \end{tabular}
      \tablenotetext{a}{
	The positions for the invisible or unkown illuminating sources of IRN I-4, II-1, III-2, 
	and IV-5 are derived by our polarization data; 
	IRN IV-2 is from \citet{krae99}, IRN IV-3 is from the Spitzer GLIMPSEII survey, 
	IRN I-2 and 3 are from \citet{debu02}, IRN V-1 and 2 are from \citet{hash06}, 
	IRN IV-6 is from \citet{bals01},
	the invisible illuminating sources of IRN II-2 cannot be well determined from 
        our data, and the other visible illuminating sources at NIR wavelengths are from 2MASS-PSC.
      }
      \tablenotetext{b}{
	Identifications are performed with IR and radio sources;
	KDJ I-3 and IV-3 are from \citet{krae99}, MM2 and 3 are from \citet{sand99},
	DPT00 2 is from \citet{debu02},
	WN-A1 and KDJ V-4 are from \citet{hash06}, NGC 6334 C and F are from \citet{rodr82},
	G351.24+0.65 is from \citet{mora90}, and the other NIR sources are from \citet{stra89}.
	``invisible'' means invisible at NIR wavelengths.
      }
      \tablenotetext{c}{
	From \citet{broo01} and \citet{krae99}.
      }
      \tablenotetext{d}{
	The $K$ band magnitude of the sources are from \citet{stra89}.
      }
      \tablenotetext{e}{
	The zero-age main-sequence (ZAMS) spectral type for the illuminating sources of
	each IRN; IRN IV-2 is from \citet{jack99}, 
	IRN I-2 is from \citet{debu02}, IRN I-3 is from \citet{pers98}, 
	IRN V-1 and 2 are from \citet{hash06}, IRN I-5, IV-1, IV-4, and V-3 are from \citet{stra89},
	IRN I-1 and III-1 are from \citet{rodr82}, and IRN IV-6 is from \citet{bals01}.
	Since the IR/radio data are not available, 
	the illuminating source of IRN I-4, II-1, 2, III-2, IV-3 and 5 are unknown.
      }
    \end{center}
  \end{table}
  
\end{document}